\documentclass[times]{aastex631}
\usepackage{amsmath}
\usepackage{booktabs}
\usepackage{bm}
\usepackage{amsfonts,times}
\usepackage{mathrsfs}
\usepackage{ulem}

\def\sunm{M_\odot}

\def\ihep{Key Laboratory for Particle Astrophysics, Institute of High Energy Physics,
Chinese Academy of Sciences, 19B Yuquan Road, Beijing 100049, China}

\def\UCASastro{School of Astronomy and Space Sciences, University of Chinese Academy of Sciences, 
19A Yuquan road, Beijing 100049, China}

\def\UCASphy{School of Physical Sciences, University of Chinese Academy of Sciences, 
19A Yuquan road, Beijing 100049, China}

\def\NAOC{National Astronomical Observatory of China, 20A Datun Road, Beijing 100020, China}

\received{XXX}
\revised{YYY}
\accepted{ZZZ}

\begin{document}

\title[Weakly warped accretion and quasar's variations]{\large Warped accretion disks and quasars with episodic periodicity of long-term variations}

\author[0009-0006-2592-7699]{Yue-Chang Peng}
\affil{\ihep}
\affil{\UCASphy}

\author[0000-0001-9449-9268]{Jian-Min Wang}
\affil{\ihep}
\affil{\UCASastro}
\affil{\NAOC}
\email{wangjm@ihep.ac.cn}

\author{Pu Du}
\affil{\ihep}

\author{Shuo Zhai}
\affil{\ihep}
\affil{\UCASphy}

\author{Yan-Rong Li}
\affil{\ihep}


\begin{abstract}
It has been found that some quasars are undergoing quasi-periodic variations (most of them with damped amplitudes) in optical bands from long-term monitoring campaigns, but how to explain the origin of such light curve variations still remains an open question. 
In this paper, we use the warped accretion disks model to explain the quasi-periodical variations. This model employs a free-bending wave traveling in an accretion disk which causes the orientation of the central part of the disk to oscillate from the line of sight, resulting in a quasi-periodical variation. We numerically solve the governing equation of warp propagation and calculate the simulated R-band light curves, finding that the periodical light curves generated by this model have damped amplitudes. To compare with observations, we select SDSSJ134820.42+194831.5 as a preliminary example from a sample of periodic quasar candidates by combining CRTS with other public survey data, and fitted its light curve with different observational angles. Our result gives a reduced $\chi^{2}\simeq 2.4$, implying that the model 
might give insights to future application of warped disk model.
\end{abstract}


\section{Introduction} 
\label{sec:intro}

Accretion disks around supermassive black holes (SMBHs) serve as the power engine for the immense radiations of active galactic nuclei (AGNs). Observations show that AGNs undergo variations across most (if not all) bands of electromagnetic spectrum, 
although the detailed physical processes in accretion disks underlying the variations are still not yet fully understood. Particularly, long-term monitoring had found a subclass of AGNs that exhibit periodical/quasi-periodical variations with periods on the timescale of 100 to 1000 days during some epochs, such as 
\citet{Graham2015} from the Catalina Real-time Transient Survey (CRTS), \citet{Charisi2016} from the Palomar Transient Factory (PTF), \citet{Liu2019} from the Pan-STARRS1 Medium 11 Deep Survey (PS1 MDS), \citet{Chen2020} from the Dark Energy Survey (DES) and the Sloan Digital Sky Survey (SDSS), and \citet{Chenyj2022} from the Zwicky Transient Facility (ZTF). While the interpretation of those (quasi-) periodicity is still under debate, there are two types of viewpoints. One is that the quasi-periodicity is purely stochastic, caused by the red-noise like variability of AGNs (e.g., \citealt{Vaughan2016}). The other one is that the quasi-periodicity might have a physical origin, in which supermassive black hole binaries are among the most popular scenarios (e.g., \citealt{Artymowicz1996, Graham2015natr, D'Orazio2015, Li2016, Li2019, Jiang2022}). However, it is found that in most cases, the periodicity is not persistent, but might disappear after several cycles, which is not easily explained by the supermassive black hole binaries hypothesis.
 
In this paper, we suggest an alternative model employing warped disks to explain the quasi-periodical variations. 
Owing to the different nature of accretion disks, warp theories can be classified into two regimes: diffusive regime \citep{Papaloizou1983, Pringle1992, Pringle1996} and wave-like regime \citep{Papaloizou1995a, Lubow2000}. The essential difference between these two theories is the different ways of angular momentum transport. For the diffusive regime, the warp is damped locally; while for the wave-like regime viscosity is insufficient to damp the warp locally, so the warp starts to propagate by pressure forces at group velocity near $c_{s}/2$ \citep{Papaloizou1995a}, where $c_{s}$ is the local sound speed. Theoretically, the condition that applies to each regime mainly depends on viscosity, under the framework of $\alpha$ disk theory,
wave-like warp theory applies to low-viscosity (or thick) disks, namely:
\begin{equation} \label{eq:1}
    \alpha<\frac{H}{R},
\end{equation}
where $\alpha$ is the viscous parameter in standard accretion disk theory, $H$ is the height of the disk, $R$ is the distance to the central supermassive black hole, and diffusive warp applies to viscid (or thin) disks. The diffusive regime is well studied both linearly and non-linearly \citep{Ogilvie1999}, while the wave-like regime is mainly studied linearly, and 
 \citet{Ogilvie2006} extended to the weakly nonlinear case. Numerical simulations 
 \citep{Lodato2010, Nixon2012} suggest that in the nonlinear case, the disk will tear into multiple planes in both regimes. Additionally, there can be parametric instability in the nonlinear case \citep{Gammie2000, Ogilvie2013}.

The disk warp may be triggered by the environment's non-coplanar perturbations, including external torques such as Lense-Thirring effect \citep{Bardeen1975}, black hole companion's tidal potential \citep{Papaloizou1995b}, and also radiation from the inner part of the accretion disk \citep{Pringle1996}. For the Lense-Thirring effect, former works suggest that both diffusive and wave-like warps will reach steady states. The diffusive regime agrees with the Bardeen-Peterson effect \citep{Scheuer1996, Martin2007}, while warp may oscillate with $R$ for the wave-like regime \citep{LOP2002, Nealon2015} . The overall precession of the disk under the Lense-Thirring effect, by generating a periodic tilt angle, are used to explain observed low frequency quasi-periodic oscillations (QPOs) in X-ray binaries \citep[such as][]{Ingram2009}. However, in the case of AGN disk, the timescale for the precession of the disk under the Lense-Thirring effect is too long to explain the observed periodic variability. 
Under the assumption that the disk precesses as a rigid pattern, we have the precessional frequency $\Omega_{\rm p}$ of the whole disk under inner torques \citep{Lodato2013, Franchini2016}:
\begin{equation}
    \Omega_{\rm p} \approx  \Omega_{\rm ext}(R_{\rm in})\left(\frac{R_{\rm out}}{R_{\rm in}} \right)^{-p-\frac{5}{2}},
\end{equation}
where $\Omega_{\rm ext}(R_{\rm in})=|\mathbf{T}|/|\mathbf{L}_{\perp}|$ is the precession frequency of the external torque at the disk inner boundary $R_{\rm in}$, $R_{\rm out}$ is the disk outer boundary, $p$ is the power-law index of the surface density of the disk. and $\mathbf{T}, \mathbf{L}_{\perp}$ are the external torque and the angular momentum component perpendicular to the direction of the torque, respectively. For the Lense-Thirring precession, we have the precession timescale:
\begin{equation}
    T_{\rm p, LT} = \frac{2\pi}{\Omega_{\rm p, LT}} \approx 75\,{\rm yr}\,\,\left(\frac{M_{\bullet}}{10^8\,M_\odot}\right) \left(\frac{a}{1.0}\right)^{-1} \left(\frac{R_{\rm in}}{6 R_{\rm g}}\right) \left(\frac{R_{\rm out}}{500 R_{\rm g}}\right)^2,
\end{equation}
with $p=-1/2$, where $M_{\bullet}$ is the mass of the central supermassive black hole, $a$ is the black hole's spin parameter, $R_{\rm g}$ is the gravitational radius. 
The same situation also goes for the circumbinary disk case \citep[][Equation (12)]{Lodato2013}. And the Pringle’s radiation instability in AGNs primarily operates on scales of $0.02$ pc, leading to much longer precession timescales \citep{Pringle1997}. Therefore, disk precession induced by these perturbations is difficult to explain the periodic variabilities we consider.



In this work, we use the wave-like warps as a model to explain some periodic variabilities. The initial warp of the circumprimary accretion disk may arise from the torque during close encounters with a black hole companion with a highly eccentric/hyperbolic orbit, and after the black hole companion moves away, the evolution of the warp can be approximated as free evolution.
This article is organized as following: Section \ref{sec:theory} introduces the basic theory of wave-like warp; Section \ref{sec:numerical cal} introduces the numerical method and results of the governing equations; Section \ref{sec:fitting} shows our results of light curve fitting. Section \ref{sec:discuss} discuss some theoretical limitations of the model; and we summarize our work in Section \ref{sec:conclusion}.

\section{Basic Equations}
\label{sec:theory}

\subsection{Warped Disk Model}

In warped disk theories, the disk is made up of concentric rings with different tilt angles. 
Assuming an initially flat accretion disk that is perturbed by a black hole companion's torque from the disk's environment, the inclination angle of each ring of the disk can align with the outside torque and finally have a distribution against radius, as we show in the discussion. 
We only consider the linear theory of wave-like warp, meaning the disk is weakly warped, so we could treat the disk evolution in a planar way while $R$ should be 3D distance in spherical coordinates.
As the outer part of the disk aligns with external torque, the bending wave starts to propagate inward, the governing equations for wave-like warp propagation are \citep{Papaloizou1995a, Lubow2000}
\begin{equation}\label{Eq.7}
    \Sigma R^{2} \Omega \frac{\partial\mathbf{l}}{\partial t} = \frac{1}{R} \frac{\partial \mathbf{G}}{\partial R} + \mathbf{T},
\end{equation}
\begin{equation}\label{Eq.6}
    \frac{\partial \mathbf{G}}{\partial t}  + \omega \mathbf{l}\times \mathbf{G} + \alpha\Omega\mathbf{G} = \frac{\Sigma H^{2} R^{3} \Omega^{3}}{4} \frac{\partial\mathbf{l}}{\partial R},
\end{equation}
where $\mathbf{l}$ is the tile vector of annulus at certain radius, $\Omega$ is the angular velocity at $R$, $\Sigma$ is the accretion disk surface density, $\textbf{G}$ is the disk's internal torque, $\textbf{T}$ is the external torque on the disk, $\omega = (\Omega^{2}-\kappa^{2})/{2 \Omega}$ is the apsidal precession frequency and $\kappa$ is the epicyclic frequency. 
We consider the disk to be Keplerian ($\omega\simeq0$), and after the initial perturbation (as we show in discussion), the warp evolves free of external torque, thus the second term on the left-hand side of Equation~(\ref{Eq.6}) and the last term on the right-hand side of Equation~(\ref{Eq.7}) vanish.
We consider quantities in Cartesian coordinates and the origin is located at the central supermassive black hole, the warp of the disk is small enough that we assume it lies on the coordinate surface.


By solving the governing equations as IVP (initial value problem), we can calculate the time evolution of the tilt vector $\mathbf{l}$. Given a certain time, the projection from the tilt vector of the annulus as each radius to the line of sight could be derived. For an optically thick disk, the radiation intensity is $I_{\lambda}=B_{\lambda}$, where $B_{\lambda}$ is the Black-body radiance. Given a temperature distribution with $R$, we can calculate the black body radiance from each radius and thus derive the radiation flux density of a certain wavelength as:
\begin{equation}\label{Eq.8}
    f_{\mathrm{\lambda}}=\int_{R_\mathrm{in}}^{R_\mathrm{out}} \mathrm{d}R\times2\pi R B_{\lambda}(\lambda,T(R))\left [ \hat{n}\cdot\mathbf{l}(t,R) \right ],
\end{equation}
where $\mathbf{l}(t,R)$ is the tilt vector at time $t$ and radius $R$, $\hat{n}$ is the line-of-sight unit vector. To compare with the real light curves, we calculate the R-band light curve using our model. In standard disk theory, the effective temperature distribution is $T_{\mathrm{eff}}^{4}=3GM_\bullet\dot{\mathcal{M}}/{8\pi \sigma_{\mathrm{SB}}R^{3}}$, where $\dot{\mathcal{M}}$ is the accretion rate and $\sigma_{\mathrm{SB}}$ is the Stefan–Boltzmann constant. However, when considering that the warping of the disk may cause the outer parts of the disk to be partially irradiated by the inner regions, and that the warping may lead to deviations from the standard theory, we generalize effective temperature to be a power-law distribution \citep{mineshige1994, kato-book}, 
%
\begin{equation}
    T_{\mathrm{eff}}=T_{\mathrm{in}}\left(\frac{R}{R_{\rm in}}\right)^{-q},
\end{equation}
here $T_{\mathrm{in}}$ and $q$ are free parameters. 
Still, the self-shadowing effect on the final observed flux due to disk warping is weak, since the linear warp theory holds when the tilts are small. We don't consider the vertical structure of the accretion disk and take $\alpha$ as a constant throughout the disk evolution, although this may not be true as the probable nature of viscosity is from turbulence, which is dependent on the disk's vertical structure. 

\subsection{Initial and Boundary Conditions}

The disk's tilt vector is described by warp and twist angle as:
\begin{gather}\label{eq_tilt}
    l_{x}=\sin\beta \cos\gamma, \\
    l_{y}=\sin\beta \sin\gamma, \\
    l_{z}=\cos\beta,
\end{gather}
where $\beta$ is the tilt angle, and $\gamma$ is the twist angle which we set to zero when calculating the case of free warps, since we find it trivial to the final results. Once the torque on the accretion disk diminishes, the disk is almost undergoing free warp evolution with an initial condition. To model this effect, we may put the initial condition after perturbation in a more general mathematical form \citep{Martin2019}:
\begin{gather}\label{eq:initial}
    \beta(0,R)=\beta_{0}\left[ \frac{1}{2}\tanh\left ( \frac{R-R_{\mathrm{warp}}}{R_{\mathrm{width}}} \right )+\frac{1}{2} \right], \\
    \gamma(0,R)=\gamma_{0}\left [ \frac{1}{2}\tanh\left ( \frac{R-R_{\mathrm{warp}}}{R_{\mathrm{width}}} \right )+\frac{1}{2} \right ].
\end{gather}
We shall note that the exact values of $\beta_{0}$, $R_{\mathrm{warp}}$ and $R_{\mathrm{width}}$ depend on the disk's properties and the strength of perturbation. We choose them to be variable parameters during fitting with observations, while fixed when exploring parameter space.

To solve the partial differential equations, we assume the surface density
\begin{equation}\label{}
    \Sigma=\Sigma_{\rm in}\left ( \frac{R}{R_{\mathrm{in}}} \right )^{-p},
\end{equation}
to be a power-law distribution through the disk. We note the final results are independent on $\Sigma_{\rm in}$ in Equation~(\ref{Eq.11}) and Equation~(\ref{Eq.12}). we take $\mathbf{G}, {\partial \mathbf{l}}/{\partial R}=0$ at the disk boundary, meaning free of inertial torque.  We take $p$, $q$, $R_{\mathrm{out}}$, $T_{\mathrm{in}}$ $\beta_{\mathrm{0}}$ and $\alpha$ as free parameters in our model. Once these parameters and boundary conditions are given, we can solve the evolution equations and finally calculate the luminosity for a given source.

\section{Numerical Calculation}
\label{sec:numerical cal}

\subsection{Method}

We use the linear analytic model with the one-dimensional numerical method described by \cite{Martin2019} to get a freely propagating bending wave. We solve Equation~(\ref{Eq.7}) and Equation~(\ref{Eq.6}) using the first-order finite difference method with a forward-time-central-space (FTCS) lattice. We follow the steps described in \cite{Facchini2013} to obtain the dimensionless form. 
We set $t=R_{\rm in}(\Omega_{\rm in}H_{\mathrm{in}})^{-1}\tau$, $R=R_{\rm g}r$, $\Omega=\Omega_{0}r^{-3/2}$,
$\mathbf{G} = \Sigma_{\rm in}R_{\rm g}^{3}\Omega_{0}^{2} H_{\rm in} \mathbf{g}$ where $\tau$ is the dimensionless time, $r$ is the dimensionless radius, $\mathbf{g}$ is the dimensionless internal torque, quantities with subscript \textbf{in} denote the values at the disk's inner boundary $R_{\rm in}$ which is fixed to $6R_{\mathrm{g}}$, and $\Omega_0$ is the angular frequency at $R_{\rm g}$.
We set $H=\eta_{\mathrm{in}}R_{\rm in}h^{s}$, and fix $\eta_{\mathrm{in}}=H_{\rm in}/R_{\rm in}=0.1$ throughout the work. Considering that in our model $R_{\mathrm{out}}\sim 500R_{\mathrm{g}}$ mainly reside in the middle region of Shakura-Sunyaev disks with relation $H\sim R^{21/20}\sim R$, we approximate $H \sim R$, i.e. $s=1$. 
The mass of the central supermassive black hole is taken as $M_{\bullet}=M_{8} \times 10^{8}M_{\odot}$ where $M_{\odot}$ is the solar mass. 
We don't consider the time evolution of surface density $\Sigma$ for simplicity. 

So, the dimensionless torque free form of Equation~(\ref{Eq.7}) and Equation~(\ref{Eq.6}) writes:
\begin{equation}\label{Eq.11}
    \frac{\Omega_{\rm in}}{\Omega_0r_{\rm in}}\frac{\partial \mathbf{l}}{\partial \tau}=\frac{1}{\sigma r^{3/2}}\frac{\partial \mathbf{g}}{\partial r},
\end{equation}
\begin{equation}\label{Eq.12}
    \frac{\Omega_{\rm in}}{\Omega_0r_{\rm in}}\frac{\partial \mathbf{g}}{\partial \tau} = \frac{\sigma h^{2}r^{-3/2}}{4}\frac{\partial \mathbf{l}}{\partial r}-\alpha r^{-\frac{3}{2}}\frac{1}{\eta_{\rm in}r_{\rm in}} \mathbf{g},
\end{equation}
where $r_{\rm in}=R_{\rm in}/R_{\rm g}$.
These equations are solved in Cartesian coordinate with $x$, $y$, and $z$ components as an initial condition problem. 


\subsection{Results} 
\label{sec:results}

\subsubsection{General Properties}
To get a general overview of warp evolution, we can consider Equation~(\ref{Eq.11}) and Equation~(\ref{Eq.12}) as a one-dimensional problem. If we ignore the damping term and use the WKB approximation, we can get a single equation for tilt angle $l$: 
\begin{equation}
    \frac{\partial ^2 l}{\partial \tau^2}=\frac{1}{4} \frac{h^2}{r^3} \frac{\partial ^2 l}{\partial r^2} \simeq \frac{1}{4r} \frac{\partial ^2 l}{\partial r^2},
\end{equation}
which is the wave equation. We can separate the solution as $l=f(t)U(r)$ and use the Neumann boundary condition to get the wave solution as:
\begin{equation}
    l \sim \sum_{n}\left[A_n\cos(n\frac{h}{2\sqrt{r^3}}t)+B_n\sin(n\frac{h}{2\sqrt{r^3}}t) \right] \times \left( C_n\cos nr+D_n\sin nr \right ),
\end{equation}
where $n$ is integer and $A_n, B_n, C_n, D_n$ are constants depending on the boundary and initial conditions. We can see $l$ oscillating in time.

\begin{figure*}
    \centering
    \includegraphics[width=1.0\textwidth]{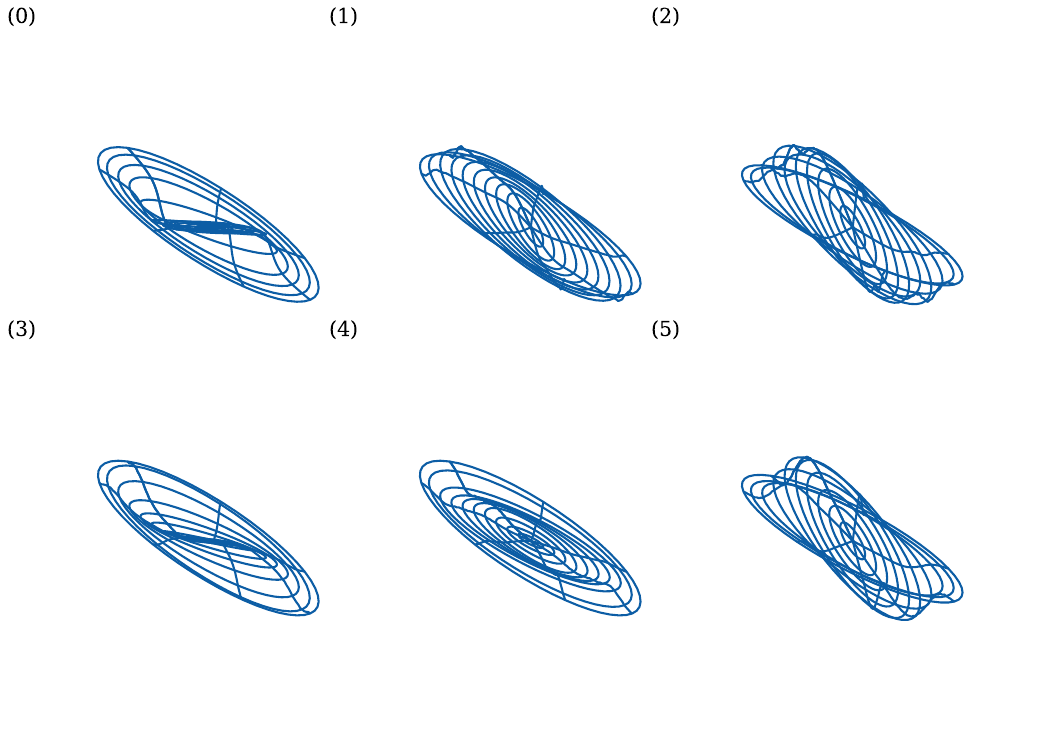}
    \caption{3D disk shapes showing warp evolution on the disk with time increases from $(0)$ to $(5)$. The central black hole is located at the center of the mesh, and the black hole mass is set to be $10^{8}\sunm$. Figures are taken every 500 days. We can see the disk's orientation is oscillating. For the purpose of displaying the distinction, the tilts of the disk shown in the figure are exaggerated.}
    \label{fig:evolution}
\end{figure*}

\begin{figure*}
    \centering
    \includegraphics[width=1.0\textwidth]{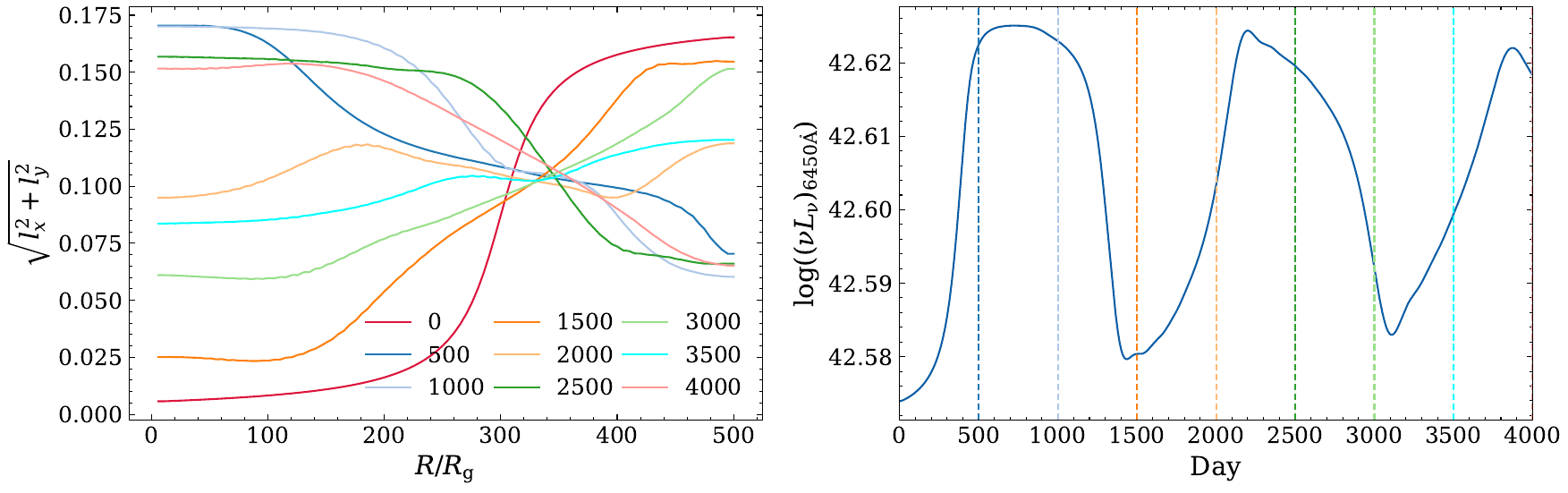}
    \caption{Left panel shows the time evolution of disk tile angle $\beta$ against dimensionless radius $r$ with time step same as Fig.~\ref{fig:evolution}, right panel shows the integrated light curve in R-band luminosity $(\nu L_\mathrm{\nu})_\mathrm{6450\mathring{A}}$ using our model. }
    \label{fig:fig2}
\end{figure*}

We show an example of warped disk evolution in Fig.~\ref{fig:evolution} and Fig.~\ref{fig:fig2} directly solved from Equation~(\ref{Eq.11}) and Equation~(\ref{Eq.12}), 
The initial warp is generated at the outer boundary of the disk by perturbations, and then propagates as a wave between the inner part and outer part of the disk. This process causes the tilt vector of the inner part of the disk to oscillate, As shown in Fig.~\ref{fig:fig2}. 

\subsubsection{Parameter Dependence}

\begin{table*}
    \centering
    \caption{Parameters for comparison}
    \label{tab:table1}
    \begin{tabular}{cccccccccc}
    \hline
    $R_{\mathrm{out}}/R_{\mathrm{g}}$ & $T_{\mathrm{in}}/K$ & $M_{8}$ & $p$ & $q$ & $\alpha$ & $R_{\mathrm{warp}}/R_{\mathrm{out}}$ & $R_{\mathrm{width}}/R_{\mathrm{out}}$ & $\beta_{0}$ & $H_{\mathrm{in}}/R_{\mathrm{in}}$ \\
    \hline
    500 & 40000 & 1 & 0.5 & 0.75 & 0.01 & 0.5 & 0.2 & $10^{\circ}$ & 0.1\\
    \hline
    \end{tabular}
\end{table*}

\begin{figure*}
    \centering
    \includegraphics[width=1.0\textwidth]{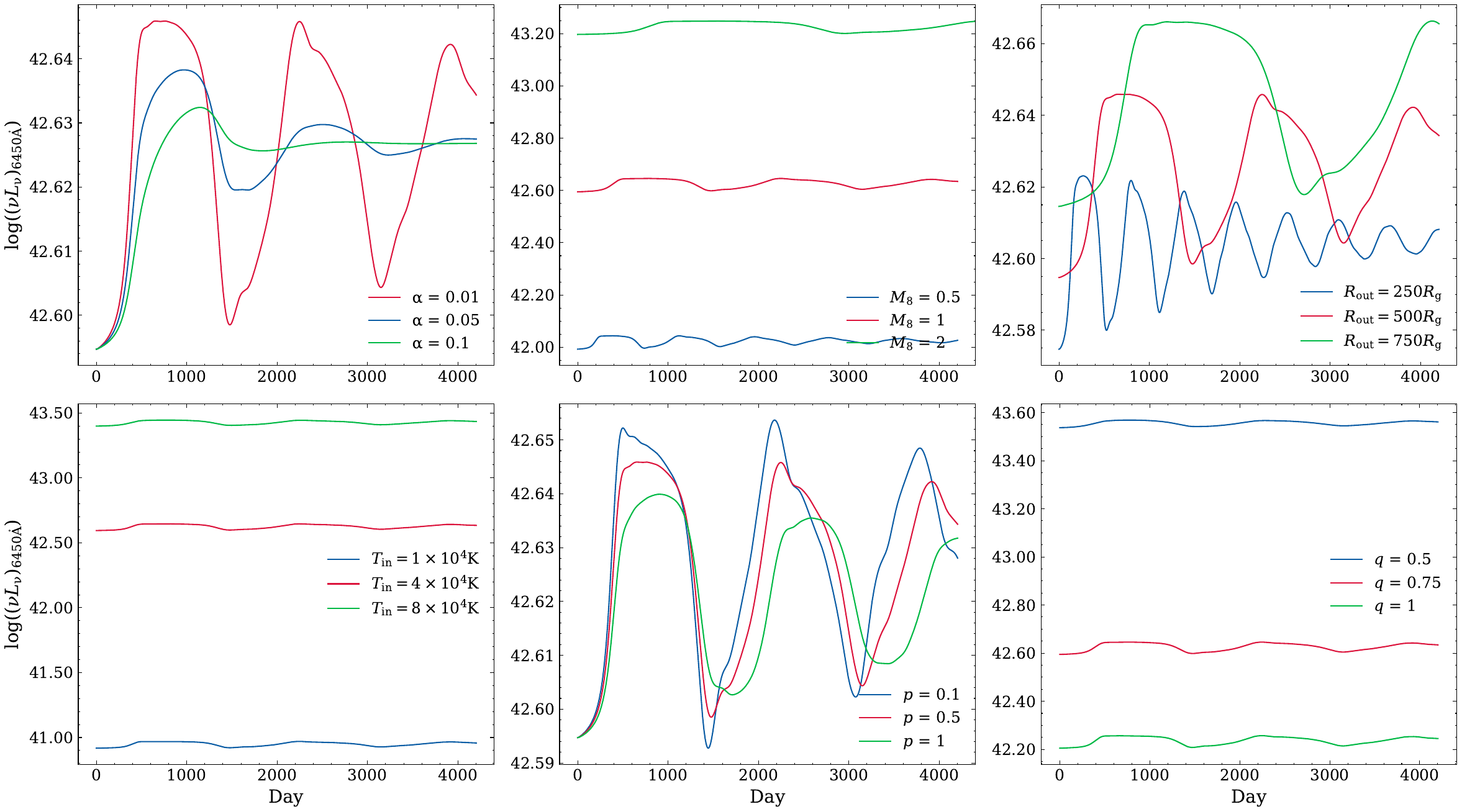}
    \caption{The simulated light curve with different parameters in our model. In each subplot we choose different $\alpha$, $M_{8}$, $R_{\mathrm{out}}$, $T_{\mathrm{in}}$, $p$, $q$ respectively, while other parameters remains to be the value in Table~\ref{tab:table1}. We highlight the cases where the parameter values all from Table~\ref{tab:table1} with crimson lines.}
    \label{fig:parameters}
\end{figure*}

We explore the light curve dependence mainly on these dimensionless parameters: $R_{\mathrm{out}}$, $M_{8}$, $\alpha$, $p$, $q$ and $T_{\mathrm{in}}$. We define the value of parameters in Table~\ref{tab:table1} and explore each parameter dependence with other parameters fixed. We have some flexibility in choosing the values of these parameters within reasonable ranges, such as $q=0.75$ is the value in Shakura-Sunyaev disks, $T_{\rm in}=40000\rm K$ is approximately the inner radius temperature of Shakura-Sunyaev disks at accretion rate $\dot{\mathcal{M}}=0.1\dot{\mathcal{M}}_{\rm Edd}$ with supermassive black hole mass $M_8=1$, the value of $p$ is somewhat arbitrary, but it is comparable to the Shakura-Sunyaev disks where $p=0.6$ in the middle region. The redshift was set to $z=0$ for parameters comparison. We show our results from Fig.~\ref{fig:parameters}

As shown in Fig.~\ref{fig:parameters}, $\alpha$ mainly affects the damping of the periodic light curve's amplitude. A larger $\alpha$ corresponds to a larger damp. The physical nature is that as the $\alpha$ parameter stands for the viscosity of the disk, a more viscid disk damps the propagating bending wave more efficiently. This property is due to the damping term in Equation~(\ref{Eq.6}), which is proportional to $\alpha \Omega$.
%
%
The mass of the central black hole mainly affects the luminosity and period of the light curve. The more massive the black hole is, the brighter the AGN source is. This is because the gravitational radius is larger for a more massive black hole, so the overall surface area of the accretion disk is larger.
And since the period of the light curve is the turn-over time of the bending wave propagating throughout the disk, the period of the light curve is also longer.
We can calculate the turn-over timescale as the propagating speed $\sim c_{\rm s}$: 
\begin{equation}
    \tau\sim - \int_{R_{\mathrm{out}}}^{R_{\mathrm{in}}} \frac{\mathrm{d}R}{c_{s}}\sim - \int_{R_{\mathrm{out}}}^{R_{\mathrm{in}}} \frac{\mathrm{d}R}{\Omega H}.
\end{equation}
For a Keplerian disk,
the integral is:
\begin{equation}
    \tau \sim \frac{1}{\sqrt{GM_{\bullet}}}R_{\rm g}^{\frac{3}{2}} \sim\frac{GM_{\bullet}}{c^{3}}.
\end{equation}
So, a more massive black hole corresponds to a longer propagating timescale, meaning a longer period.
%
%
Same as the $M_{8}$ case, a larger disk has a longer timescale and a higher luminosity.
%
%
The dependence of period timescale on  $p$ compared with $M_{8}$ and $R_{\mathrm{out}}$ is a much weaker, but trends of these light curves are different. This is because with different $p$ the outer part of the disk takes different portion of the disk's total mass, so when the outer part of the disk aligns with the external perturbation with the same tilt angle, the final angle of the disk after the warp evolution damped is different. We should address that we do not consider the time evolution of surface density, which is not reasonable, as warp evolution shall take effect on the surface density distribution and vice versa. But we don't think there will be significant effects on periodic variability.
%
%
For the $q$ dependence in Fig.~\ref{fig:parameters}, the smaller the $q$ is, the flatter the temperature distribution is, which means the outer part of the disk has the same radiation capability as the inner part, so the total radiation is larger with smaller $q$, and the period is inconspicuous for there is no significant area where the disk emits the most.

\section{Comparison with observations}
\label{sec:fitting}
\begin{table*}
    \centering
    \caption{Physical Property of Selected Source}
    \label{tab:selected source}
    \begin{tabular}{ccccc}
    \hline
        \textbf{Name} & \textbf{Observed Period} & \textbf{$\log_{10}M/\sunm$} & \textbf{Redshift} & \textbf{Rest-Frame Period} \\ 
        \textbf{} & (days) & & \textbf{} & (days) \\
    \hline
        $\mathrm{SDSSJ}134820.42+194831.5$ & $1482$ & $7.63$ & $0.594$ & $930$ \\
    \hline
    \end{tabular}
\end{table*}

To compare with observed periodic variability, we first selected the quasi-periodic variable sources from the sample identified by \cite{Graham2015a}, they have identified a sample of $111$ periodical quasar candidates using data from Catalina Real-time Transient Survey (CRTS). We then extend the temporal range of these candidates' light curves to about 20 years, by using public survey data from ZTF (2017-present), DES (2013-2019), PS1 (2009–2014), PTF (2009–2012) and ASAS-SN (2012-present), and the R-band filter is chosen based on PS1 or Catalina. We revisit $81$ out of $111$ candidates with generally more than $3$ period cycles. 

Since the theoretical model requires the periods of the light curves to remain almost the same, while the amplitude of oscillation is damped, we reanalyzed the periodicity of the candidates data using the generalized Lomb–Scargle (GLS) periodogram, employing a sine-like function with varying amplitude to get their observed periods:
\begin{equation}
    \mathrm{Mag} = A(1+\dot{A}t)\sin{\left(\frac{2\pi t}{P_{\rm obs}}+\phi \right)} + M_0,
\end{equation}
where $A$, $\dot{A}$, $P$, $\phi$ and $M_0$ are light curve's amplitude, changing rate of amplitude, period, phase and offset, respectively. More thorough and detailed analysis will be presented in a forthcoming work by Zhai et al (in preparation). We therefore selected $\mathrm{SDSSJ}134820.42+194831.5$ as a preliminary example, which satisfies our requirement based on its properties. 



Other physical parameters of the source, such as black hole mass and redshift, were derived from \cite{Graham2015} and \cite{Rakshit2020} who have conducted 520,000 quasars in the ZTF catalog. Once dimensionless black hole mass $M_{8}$ and redshift $z$ are given, we use $R_{\mathrm{out}}/R_{\mathrm{g}}$, $T_{\mathrm{in}}/\rm K$, $p$, $q$, $\alpha$, $s$, $R_{\mathrm{warp}}/R_{\mathrm{out}}$, $R_{\mathrm{width}}/R_{\mathrm{out}}$ and $\beta_{\mathrm{0}}$ as free parameters to fit the light curve. Radiation intensity derived directly from our model is in the rest frame and should be transformed into the observer's frame. We transform the observed wavelength and period to the accretion disk's rest-frame using $\lambda_{\mathrm{rest}}=\lambda_{\mathrm{obs}}/(1+z)$ and $P_{\mathrm{rest}}=P_{\mathrm{obs}}/(1+z)$. We show the observed period by GLS periodogram and the corresponding rest-frame period in Table~\ref{tab:selected source}. To get the magnitude of the source, we calculate the luminosity distance by using standard $\Lambda\mathrm{CDM}$ cosmology model $\left (H=70\mathrm{km/s/Mpc}, \Omega_{\rm M}=0.3 \right)$, and calculate the AB magnitude by using flux in observational frame.

\begin{figure*}
    \centering
    \includegraphics[width=1.0\textwidth]{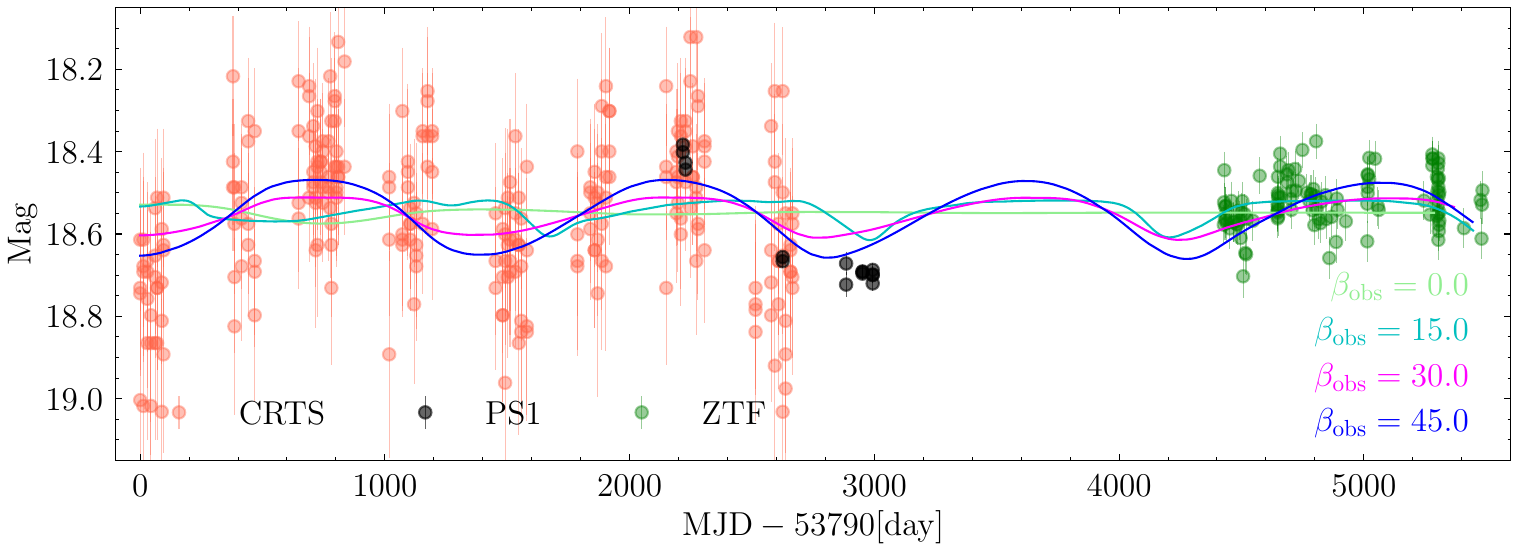}
    \caption{Fit of SDSSJ134820.42+194831.5 with multiple $\beta_{\mathrm{obs}}$. We fit the disk with different observation angles, for it's a purely geometrical effect. The fitting results are shown in the Table~\ref{tab:table3}}
    \label{fig:fig8}
\end{figure*}

We then use the Python package {\tt lmfit} \citep{Newville2016} to fit our model with these quasi-periodic light curves. Because of the uncertainty of the observational angle to the disk, we choose it as a fixed parameter when we fit the light curve with different $\beta_{\mathrm{obs}}$ from $0^{\circ}$(face on) to $45^{\circ}$(continuum obscured by AGN's dust torus). To compare the results of different observational angle, we show our results of SDSSJ134820.42+194831.5 in Fig.~\ref{fig:fig8}. The fitting parameters and reduced $\chi^{2}$ are show in Table~\ref{tab:table3}. 
Since we set the maximum value of $\beta_0$ to $20^{\circ}$, the fitted $\beta_0$ values were mostly close to this upper limit.
We find our best reduced $\chi^2$ roughly to be $2.4$,
and these parameters in our model are highly degenerated and show a wide range of parameter space with almost the same reduced $\chi^{2}$.

\begin{table*}
    \centering
    \caption{Fitting Parameters of SDSSJ134820.42+194831.5}
    \label{tab:table3}
    \begin{tabular}{lccccccccccc}
        \hline
        \textbf{$\beta_{\mathrm{obs}}$} & \textbf{$R_{\mathrm{out}}$} & 
        \textbf{$T_{\mathrm{in}}$} & \textbf{$\alpha$} & \textbf{$\beta_{\mathrm{0}}$} & 
        \textbf{$p$} & \textbf{$q$} & $R_{\mathrm{warp}}/R_{\mathrm{out}}$ & $R_{\mathrm{width}}/R_{\mathrm{out}}$ & $H_{\mathrm{in}}/R_{\mathrm{in}}$ & $s$ & \textbf{$\mathrm{Reduced}-\chi^{2}$} \\ 
        $(^\circ)$ & $(R_\mathrm{g})$ & $(10^4\mathrm{K})$ & & $(^\circ)$ \\
        \hline
        $0$ & $609$ & $132$ & $0.001$ & $19.9$ & $0.50$ & $0.99$ & $0.63$ & $0.18$ & $0.11$ & $0.96$ & $3.66$\\
        $15$ & $599$ & $97$ & $0.01$ & $19.9$ & $0.58$ & $0.91$ & $0.85$ & $0.22$ & $0.11$ & $0.91$ & $4.08$ \\
        $30$ & $396$ & $137$ & $0.001$ & $19.9$ & $0.68$ & $1.00$ & $0.68$ & $0.15$ & $0.11$ & $0.92$ & $2.97$ \\
        $45$ & $626$ & $99$ & $0.001$ & $19.9$ & $0.59$ & $0.88$ & $0.65$ & $0.20$ & $0.11$ & $0.92$ & $2.47$\\
        \hline
    \end{tabular}
\end{table*}





\section{Discussion} 
\label{sec:discuss}

\subsection{Relation between $\alpha$ and $H/R$}

Theoretically, wave-like warp evolution mainly applies to the disk with $\alpha< {H}/{R}$, and the disk's viscosity is supposedly contributed from turbulence viscosity, described by the $\alpha$ parameter. Analytical works suggest magnetorotational instability (MRI) may be the underlying cause of turbulence \citep{Balbus1991, Hawley1994}, and hydrodynamical simulations of MRI turbulence indicate the disk's $\alpha$ parameter to be $0.02 \sim 0.03$ \citep{Simon2012}. Early numerical simulations of bending waves indicate $\alpha \sim 0.04$ \citep{Nelson1999}, which value is comparable to disk's aspect ratio $H/R$ numerically solved directly by standard $\alpha$ disk theory. We also note that for wave-like regime, when the $\alpha$ parameter is large 
it can also represent the similar behavior of diffusive warps \citep{Nixon2016, Facchini2013}, so the consideration of wave-like warps may be acceptable.


In comparison with observations, to get a smaller reduced $\chi^2$, we find that $\alpha$ tends to be as small as possible, and $\eta_{\mathrm{in}}$ tends to be as large as possible, which is self-consistent with the wave-like regime.

\subsection{Effective Temperature Profile}
From an observational perspective, the effective temperature can be considered as a power-law distribution, with the power-law index $q$ ranging from $0.5-0.75$ \citep{mineshige1994, Cheng2019}. And theoretical models of accretion disks \citep{kato-book} suggest standard accretion disk $p=3/4$ and irradiated disk $p\sim3/7$, with varying power-law indexes. When the accretion disk is warped, the deviation of the power-law index from disk models may be caused by the outer part of the disk being irradiated by the central part of the disk, and the disk's dynamics being changed at the warp radius.

\subsection{Disk Size}

There are a few constraints on the disk size, a condition which should be satisfied is that the outer boundary resides in the self-gravity (SG) radius where self-gravitation takes effect and the disk may suffer instabilities and finally break into clouds, giving the origin of broad line region (BLR). \cite{Goodman2003} has approximated the SG radius $R_{\mathrm{SG}}/R_{\mathrm{g}}\simeq3.1\times10^{3}\alpha_{0.1}^{2/9}\epsilon_{\mathrm{Edd}}^{4/9}M_{8}^{-2/9}$, where $\alpha_{0.1}=\alpha/0.1$, $\epsilon_{\mathrm{Edd}}$ is the Eddington ratio. 
For SDSSJ134820.42+194831.5 with $M_{8}=0.43$ and $\epsilon_{\mathrm{Edd}}\simeq(\nu L_\mathrm{\nu})_\mathrm{5100\mathring{A}}/L_{\mathrm{Edd}} \simeq0.28$, we calculate the SG radius to be $R_{\mathrm{SG}}/R_{\mathrm{g}}\sim 760$ to $1200$ with $\alpha$ ranges from $0.001$ to $0.01$, which is larger than $R_{\mathrm{out}}$ in our model fitting. 



\subsection{Alignment with black hole companion}

When considering a binary black hole companion outside the disk, the external torque in Equation~\ref{Eq.7} should be considered. According to \citep{Wijers1999, Lubow2000}, the external torque due to black hole companion is:
\begin{equation}
    \mathbf{T} = \frac{GM_2}{2R_{b}^4}b^{(1)}_{3/2}\left(\frac{R}{R_{b}}\right)\Sigma R\left(\mathbf{R}_{b}\cdot \mathbf{l}\right)\left(\mathbf{R}_{b}\times \mathbf{l}\right),
\end{equation}
where $M_2$ is the mass of the black hole companion, $\mathbf{R}_{\rm b}$ is the vector pointing the black hole companion, $b^{(1)}_{3/2}$ is the Laplace coefficient:
\begin{equation}
    b^{(1)}_{3/2}(x) = \frac{2}{\pi}\int_{0}^{\pi}\frac{\cos\phi \mathrm{d}\phi}{\left(1+x^2-2x\cos\phi \right)^{\frac{3}{2}}},
\end{equation}
For $x<1$, we can use the asymptotic form to simplify the Laplace coefficient \citep{Brouwer1961}:
\begin{equation}
    b^{(1)}_{3/2}(x) \simeq 3x\left(1+\frac{15}{8}x^2+\cdots \right).
\end{equation}
To investigate the perturbation of an accretion disk by a black hole companion with high eccentricity, it may be instructive to examine the case a companion with a circular orbit (and the full consideration with eccentricity will be considered in a future work). Assuming that $\mathbf{R}_{\rm b}$ lies on the $XZ$-plane at the initial time, and the inclination angle of the orbit plane of the black hole companion is $\beta_{\rm b}$, thus $\mathbf{R}_{\rm b}$ can be expressed as:
\begin{equation}
    \mathbf{R}_{\rm b} = R_{\rm b} 
    \begin{pmatrix}
        \cos\Omega_{\rm b}t \cos\beta_{\rm b}\\
        \sin\Omega_{\rm b}t \\
        \cos\Omega_{\rm b}t \sin\beta_{\rm b}
    \end{pmatrix},
\end{equation}
where $\Omega_{\rm b}$ is the angular velocity of the black hole companion:
\begin{equation}
    \Omega_{\rm b} = \sqrt{\frac{q_{\rm b}}{1+q_{\rm b}}}\Omega_0 \left(\frac{R_{\rm b}}{R_{\rm g}}\right)^{-\frac{3}{2}},
\end{equation}
where $q_{\rm b}=M_2/M_{\bullet}$ is the mass ratio. Combined with above equations, we can rewrite the external torque density as:
\begin{equation}
    \mathbf{T} = \frac{3}{2}\frac{GM_2R^2\Sigma}{R_{\rm b}^3}\mathbf{e}_{\rm b},
\end{equation}
where
\begin{equation}
    \mathbf{e}_{\rm b} = \left(l_{x}\cos\Omega_{\rm b}t \cos\beta_{\rm b}+l_{y}\sin\Omega_{\rm b}t+l_{z}\cos\Omega_{\rm b}t \sin\beta_{\rm b} \right)  \times
    \begin{pmatrix}
        l_{z}\sin\Omega_{\rm b}t -l_{y} \cos\Omega_{\rm b}t \sin\beta_{\rm b}  \\
        l_{x}\cos\Omega_{\rm b}t \sin\beta_{\rm b}  - l_{z}\cos\Omega_{\rm b}t \cos\beta_{b}  \\
        l_{y}\cos\Omega_{\rm b}t \cos\beta_{\rm b}  - l_{x}\sin\Omega_{\rm b}t 
    \end{pmatrix},
\end{equation}
We should note that the centrifugal force has no net torque on the disk. Thus, we can solve the warp evolution under the potential of a binary companion once $M_2, R_{\rm b}, \beta_{\rm b}$ are specified. We set the disk to be initially flat when solving the forced evolution under the companion black hole potential.
The dimensionless form of Equation~\ref{Eq.7} reads:
\begin{equation}
    \frac{\Omega_{\rm in}}{\Omega_0r_{\rm in}}\frac{\partial \mathbf{l}}{\partial \tau}=\frac{1}{\sigma r^{3/2}}\frac{\partial \mathbf{g}}{\partial r} + \frac{3}{2}r^{\frac{3}{2}}\frac{q_{\rm b}}{\eta_{\rm in}r_{\rm in}}\left(\frac{R_{\rm g}}{R_{\rm b}}\right)^3 \mathbf{e}_{\rm b},
\end{equation}
From the above equation, we can see that tidal torques only dominate at large radii. 
For a black hole companion with circular orbit, the outer radius of accretion disk due to tidal truncation is $R_{\rm out}\approx 0.3 R_{b}$ \citep{Lubow2000}. For disk's outer boundary $R_{\rm out}=500R_{\rm g}$ the same value as in former cases, this corresponds to the binary separation $R_{\rm b}\approx 1600R_{\rm g}$. However, for eccentric orbits, the periapsis of the companion may be smaller than the separation of the circular case. 
To numerically solve the tilt evolution under external torque, we adopt the numerical method described in \citep{LOP2002}, by using central-time-forward-space (leapfrog) lattice, and since the boundary conditions only apply to internal torque $\mathbf{G}$, we place the spatial grids of $\mathbf{l}$ and $\mathbf{G}$ alternately.

\begin{figure}
    \centering
    \includegraphics[width=0.5\linewidth]{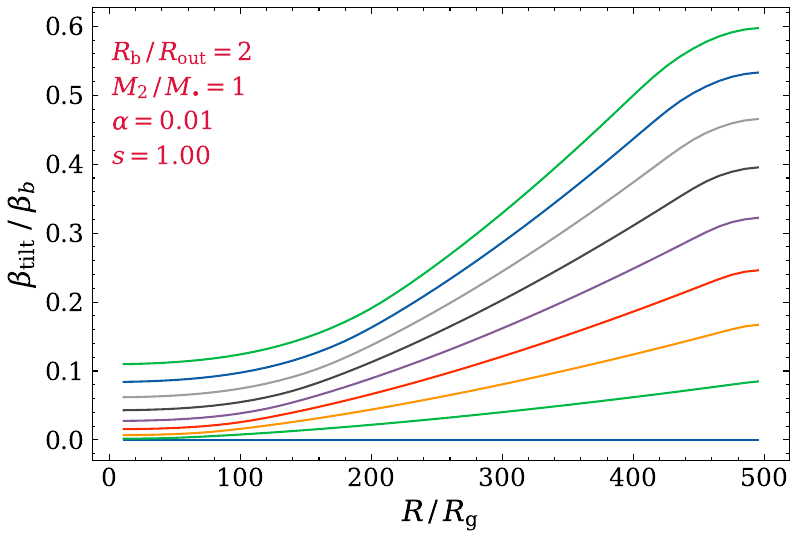}
    \caption{The tilt angle evolution for accretion disk with a black hole companion. The binary separation is set to $2R_{\rm out}$, with black hole mass ratio $q_{\rm b}=1$. The tilts are separated by $30$ days.}
    \label{fig:BHB_tilt}
\end{figure}

We show our results in Figure~\ref{fig:BHB_tilt}.
We can see that at the initial moment, the disk responds to the gravitational potential of the binary black holes, causing warping in the outer region. After the companion moves away from the accretion disk, the evolution of the warp can be approximated as free evolution with initial conditions, as described in Section~\ref{sec:theory}.


\section{Conclusion} 
\label{sec:conclusion}

In this paper, we consider the projection effect of the low-viscosity weakly warped disk as an underlying mechanism to interpret the damping periodic optical variations observed by long-term surveys e.g. Catalina, ZTF. First, we numerically solved the linear wave-like warp equations to get the warp and tilt angle evolution over time, from which we can get a warp wave turn-over time that is approximately the rest-frame period timescale. 
Secondly, we set an effective temperature profile and calculate its black-body radiation by taking the tilt angle into consideration. The observed flux is largest when the tile vector is along the observational vector, such that the disk is face-on. 
Then we explore the parameter space of the simulated light curves, and select quasars with a periodic R-band variability that could satisfy our model's property from \citep{Graham2015}. We select SDSSJ134820.42+194831.5 as a preliminary example with reduced $\chi^2\simeq 2.4$. 
Through fitting long-term light curves, a more extensive selection of sources exhibiting unchanging periods and decreasing amplitudes of periodic variability will be the focus of our next work.

We need to address that the real situation of periodic variability is complex and may contain multiple mechanisms with different trigger points. Any attempt to reason realistic periodic variability of certain sources involving accretion disk mechanism depends on luminosity, together with black hole mass and accretion rate describes the accretion nature of the source; variability timescale and amplitude, which describes the variability mechanism, and the observed wavelength, which describes the position of variability. We keep in mind that this work is a first step, and the mechanism of warped disks to explain the quasi-period variability needs further development for future works.

\begin{acknowledgments}
The authors appreciate the constructive feedback from the anonymous referee, and the discussions with members of IHEP AGN group. J.M.W. thanks the support by the National Key R\&D Program of China through grants 2020YFC2201400 by NSFC-11991050, -11991054, -12333003.
\end{acknowledgments}

\vspace{5mm}


\bibliography{ref}{}

\begin{thebibliography}{}
\expandafter\ifx\csname natexlab\endcsname\relax\def\natexlab#1{#1}\fi
\providecommand{\url}[1]{\href{#1}{#1}}
\providecommand{\dodoi}[1]{doi:~\href{http://doi.org/#1}{\nolinkurl{#1}}}
\providecommand{\doeprint}[1]{\href{http://ascl.net/#1}{\nolinkurl{http://ascl.net/#1}}}
\providecommand{\doarXiv}[1]{\href{https://arxiv.org/abs/#1}{\nolinkurl{https://arxiv.org/abs/#1}}}

\bibitem[{{Artymowicz} \& {Lubow}(1996)}]{Artymowicz1996}
{Artymowicz}, P., \& {Lubow}, S.~H. 1996, \apjl, 467, L77,
  \dodoi{10.1086/310200}

\bibitem[{{Balbus} \& {Hawley}(1991)}]{Balbus1991}
{Balbus}, S.~A., \& {Hawley}, J.~F. 1991, \apj, 376, 214,
  \dodoi{10.1086/170270}

\bibitem[{{Bardeen} \& {Petterson}(1975)}]{Bardeen1975}
{Bardeen}, J.~M., \& {Petterson}, J.~A. 1975, \apjl, 195, L65,
  \dodoi{10.1086/181711}

\bibitem[{{Brouwer} \& {Clemence}(1961)}]{Brouwer1961}
{Brouwer}, D., \& {Clemence}, G.~M. 1961, {Methods of celestial mechanics}

\bibitem[{{Charisi} {et~al.}(2016){Charisi}, {Bartos}, {Haiman},
  {Price-Whelan}, {Graham}, {Bellm}, {Laher}, \& {M{\'a}rka}}]{Charisi2016}
{Charisi}, M., {Bartos}, I., {Haiman}, Z., {et~al.} 2016, \mnras, 463, 2145,
  \dodoi{10.1093/mnras/stw1838}

\bibitem[{{Chen} {et~al.}(2020){Chen}, {Liu}, {Liao}, {Holgado}, {Guo},
  {Gruendl}, {Morganson}, {Shen}, {Zhang}, {Abbott}, {Aguena}, {Allam},
  {Avila}, {Bertin}, {Bhargava}, {Brooks}, {Burke}, {Carnero Rosell},
  {Carollo}, {Carrasco Kind}, {Carretero}, {Costanzi}, {da Costa}, {Davis}, {De
  Vicente}, {Desai}, {Diehl}, {Doel}, {Everett}, {Flaugher}, {Friedel},
  {Frieman}, {Garc{\'\i}a-Bellido}, {Gaztanaga}, {Glazebrook}, {Gruen},
  {Gutierrez}, {Hinton}, {Hollowood}, {James}, {Kim}, {Kuehn}, {Kuropatkin},
  {Lewis}, {Lidman}, {Lima}, {Maia}, {March}, {Marshall}, {Menanteau},
  {Miquel}, {Palmese}, {Paz-Chinch{\'o}n}, {Plazas}, {Sanchez}, {Schubnell},
  {Serrano}, {Sevilla-Noarbe}, {Smith}, {Suchyta}, {Swanson}, {Tarle},
  {Tucker}, {Norbert Varga}, \& {Walker}}]{Chen2020}
{Chen}, Y.-C., {Liu}, X., {Liao}, W.-T., {et~al.} 2020, \mnras, 499, 2245,
  \dodoi{10.1093/mnras/staa2957}

\bibitem[{{Chen} {et~al.}(2024){Chen}, {Zhai}, {Liu}, {Guo}, {Peng}, {Li},
  {Songsheng}, {Du}, {Hu}, \& {Wang}}]{Chenyj2022}
{Chen}, Y.-J., {Zhai}, S., {Liu}, J.-R., {et~al.} 2024, \mnras, 527, 12154,
  \dodoi{10.1093/mnras/stad3981}

\bibitem[{{Cheng} {et~al.}(2019){Cheng}, {Yuan}, {Liu}, {Breeveld}, {Jin}, \&
  {Liu}}]{Cheng2019}
{Cheng}, H., {Yuan}, W., {Liu}, H.-Y., {et~al.} 2019, \mnras, 487, 3884,
  \dodoi{10.1093/mnras/stz1532}

\bibitem[{{D'Orazio} {et~al.}(2015){D'Orazio}, {Haiman}, \&
  {Schiminovich}}]{D'Orazio2015}
{D'Orazio}, D.~J., {Haiman}, Z., \& {Schiminovich}, D. 2015, \nat, 525, 351,
  \dodoi{10.1038/nature15262}

\bibitem[{{Facchini} {et~al.}(2013){Facchini}, {Lodato}, \&
  {Price}}]{Facchini2013}
{Facchini}, S., {Lodato}, G., \& {Price}, D.~J. 2013, \mnras, 433, 2142,
  \dodoi{10.1093/mnras/stt877}

\bibitem[{Franchini {et~al.}(2015)Franchini, Lodato, \&
  Facchini}]{Franchini2016}
Franchini, A., Lodato, G., \& Facchini, S. 2015, Monthly Notices of the Royal
  Astronomical Society, 455, 1946, \dodoi{10.1093/mnras/stv2417}

\bibitem[{{Gammie} {et~al.}(2000){Gammie}, {Goodman}, \&
  {Ogilvie}}]{Gammie2000}
{Gammie}, C.~F., {Goodman}, J., \& {Ogilvie}, G.~I. 2000, \mnras, 318, 1005,
  \dodoi{10.1046/j.1365-8711.2000.03669.x}

\bibitem[{{Goodman}(2003)}]{Goodman2003}
{Goodman}, J. 2003, \mnras, 339, 937, \dodoi{10.1046/j.1365-8711.2003.06241.x}

\bibitem[{{Graham} {et~al.}(2015{\natexlab{a}}){Graham}, {Djorgovski}, {Stern},
  {Drake}, {Mahabal}, {Donalek}, {Glikman}, {Larson}, \&
  {Christensen}}]{Graham2015}
{Graham}, M.~J., {Djorgovski}, S.~G., {Stern}, D., {et~al.} 2015{\natexlab{a}},
  \mnras, 453, 1562, \dodoi{10.1093/mnras/stv1726}

\bibitem[{{Graham} {et~al.}(2015{\natexlab{b}}){Graham}, {Djorgovski}, {Stern},
  {Glikman}, {Drake}, {Mahabal}, {Donalek}, {Larson}, \&
  {Christensen}}]{Graham2015natr}
---. 2015{\natexlab{b}}, \nat, 518, 74, \dodoi{10.1038/nature14143}

\bibitem[{{Graham} {et~al.}(2015{\natexlab{c}}){Graham}, {Djorgovski}, {Stern},
  {Drake}, {Mahabal}, {Donalek}, {Glikman}, {Larson}, \&
  {Christensen}}]{Graham2015a}
---. 2015{\natexlab{c}}, \mnras, 453, 1562, \dodoi{10.1093/mnras/stv1726}

\bibitem[{{Hawley} {et~al.}(1994){Hawley}, {Gammie}, \& {Balbus}}]{Hawley1994}
{Hawley}, J.~F., {Gammie}, C.~F., \& {Balbus}, S.~A. 1994, in Astronomical
  Society of the Pacific Conference Series, Vol.~54, The Physics of Active
  Galaxies, ed. G.~V. {Bicknell}, M.~A. {Dopita}, \& P.~J. {Quinn}, 73

\bibitem[{{Ingram} {et~al.}(2009){Ingram}, {Done}, \& {Fragile}}]{Ingram2009}
{Ingram}, A., {Done}, C., \& {Fragile}, P.~C. 2009, \mnras, 397, L101,
  \dodoi{10.1111/j.1745-3933.2009.00693.x}

\bibitem[{{Jiang} {et~al.}(2022){Jiang}, {Yang}, {Wang}, {Zhu}, {Lyu}, {Dou},
  {Wang}, {Wang}, {Pan}, {Liu}, {Shu}, \& {Zheng}}]{Jiang2022}
{Jiang}, N., {Yang}, H., {Wang}, T., {et~al.} 2022, arXiv e-prints,
  arXiv:2201.11633.
\newblock \doarXiv{2201.11633}

\bibitem[{Kato {et~al.}(2008)Kato, Fukue, \& Mineshige}]{kato-book}
Kato, S., Fukue, J., \& Mineshige, S. 2008, Black-Hole Accretion Disks ---
  Towards a New Paradigm ---, 549 pages, including 12 Chapters, 9 Appendices,
  ISBN 978-4-87698-740-5, Kyoto University Press (Kyoto, Japan), 2008., -1

\bibitem[{{Li} {et~al.}(2016){Li}, {Wang}, {Ho}, {Lu}, {Qiu}, {Du}, {Hu},
  {Huang}, {Zhang}, {Wang}, \& {Bai}}]{Li2016}
{Li}, Y.-R., {Wang}, J.-M., {Ho}, L.~C., {et~al.} 2016, \apj, 822, 4,
  \dodoi{10.3847/0004-637X/822/1/4}

\bibitem[{{Li} {et~al.}(2019){Li}, {Wang}, {Zhang}, {Wang}, {Huang}, {Lu},
  {Hu}, {Du}, {Bon}, {Ho}, {Bai}, {Bian}, {Yuan}, {Winkler}, {Denissyuk},
  {Valiullin}, {Bon}, \& {Popovi{\'c}}}]{Li2019}
{Li}, Y.-R., {Wang}, J.-M., {Zhang}, Z.-X., {et~al.} 2019, \apjs, 241, 33,
  \dodoi{10.3847/1538-4365/ab0ec5}

\bibitem[{{Liu} {et~al.}(2019){Liu}, {Gezari}, {Ayers}, {Burgett}, {Chambers},
  {Hodapp}, {Huber}, {Kudritzki}, {Metcalfe}, {Tonry}, {Wainscoat}, \&
  {Waters}}]{Liu2019}
{Liu}, T., {Gezari}, S., {Ayers}, M., {et~al.} 2019, \apj, 884, 36,
  \dodoi{10.3847/1538-4357/ab40cb}

\bibitem[{{Lodato} \& {Facchini}(2013)}]{Lodato2013}
{Lodato}, G., \& {Facchini}, S. 2013, \mnras, 433, 2157,
  \dodoi{10.1093/mnras/stt878}

\bibitem[{{Lodato} \& {Price}(2010)}]{Lodato2010}
{Lodato}, G., \& {Price}, D.~J. 2010, \mnras, 405, 1212,
  \dodoi{10.1111/j.1365-2966.2010.16526.x}

\bibitem[{{Lubow} \& {Ogilvie}(2000)}]{Lubow2000}
{Lubow}, S.~H., \& {Ogilvie}, G.~I. 2000, \apj, 538, 326,
  \dodoi{10.1086/309101}

\bibitem[{{Lubow} {et~al.}(2002){Lubow}, {Ogilvie}, \& {Pringle}}]{LOP2002}
{Lubow}, S.~H., {Ogilvie}, G.~I., \& {Pringle}, J.~E. 2002, \mnras, 337, 706,
  \dodoi{10.1046/j.1365-8711.2002.05949.x}

\bibitem[{{Martin} {et~al.}(2007){Martin}, {Pringle}, \& {Tout}}]{Martin2007}
{Martin}, R.~G., {Pringle}, J.~E., \& {Tout}, C.~A. 2007, \mnras, 381, 1617,
  \dodoi{10.1111/j.1365-2966.2007.12349.x}

\bibitem[{{Martin} {et~al.}(2019){Martin}, {Lubow}, {Pringle}, {Franchini},
  {Zhu}, {Lepp}, {Nealon}, {Nixon}, \& {Vallet}}]{Martin2019}
{Martin}, R.~G., {Lubow}, S.~H., {Pringle}, J.~E., {et~al.} 2019, \apj, 875, 5,
  \dodoi{10.3847/1538-4357/ab0bb7}

\bibitem[{{Mineshige} {et~al.}(1994){Mineshige}, {Hirano}, {Kitamoto},
  {Yamada}, \& {Fukue}}]{mineshige1994}
{Mineshige}, S., {Hirano}, A., {Kitamoto}, S., {Yamada}, T.~T., \& {Fukue}, J.
  1994, \apj, 426, 308, \dodoi{10.1086/174065}

\bibitem[{{Nealon} {et~al.}(2015){Nealon}, {Price}, \& {Nixon}}]{Nealon2015}
{Nealon}, R., {Price}, D.~J., \& {Nixon}, C.~J. 2015, \mnras, 448, 1526,
  \dodoi{10.1093/mnras/stv014}

\bibitem[{{Nelson} \& {Papaloizou}(1999)}]{Nelson1999}
{Nelson}, R.~P., \& {Papaloizou}, J. C.~B. 1999, \mnras, 309, 929,
  \dodoi{10.1046/j.1365-8711.1999.02894.x}

\bibitem[{{Newville} {et~al.}(2016){Newville}, {Stensitzki}, {Allen}, {Rawlik},
  {Ingargiola}, \& {Nelson}}]{Newville2016}
{Newville}, M., {Stensitzki}, T., {Allen}, D.~B., {et~al.} 2016, {Lmfit:
  Non-Linear Least-Square Minimization and Curve-Fitting for Python},
  Astrophysics Source Code Library, record ascl:1606.014.
\newblock \doeprint{1606.014}

\bibitem[{{Nixon} \& {King}(2016)}]{Nixon2016}
{Nixon}, C., \& {King}, A. 2016, in Lecture Notes in Physics, Berlin Springer
  Verlag, ed. F.~{Haardt}, V.~{Gorini}, U.~{Moschella}, A.~{Treves}, \&
  M.~{Colpi}, Vol. 905, 45, \dodoi{10.1007/978-3-319-19416-5\_2}

\bibitem[{{Nixon} \& {King}(2012)}]{Nixon2012}
{Nixon}, C.~J., \& {King}, A.~R. 2012, \mnras, 421, 1201,
  \dodoi{10.1111/j.1365-2966.2011.20377.x}

\bibitem[{{Ogilvie}(1999)}]{Ogilvie1999}
{Ogilvie}, G.~I. 1999, \mnras, 304, 557,
  \dodoi{10.1046/j.1365-8711.1999.02340.x}

\bibitem[{{Ogilvie}(2006)}]{Ogilvie2006}
---. 2006, \mnras, 365, 977, \dodoi{10.1111/j.1365-2966.2005.09776.x}

\bibitem[{{Ogilvie} \& {Latter}(2013)}]{Ogilvie2013}
{Ogilvie}, G.~I., \& {Latter}, H.~N. 2013, \mnras, 433, 2420,
  \dodoi{10.1093/mnras/stt917}

\bibitem[{{Papaloizou} \& {Lin}(1995)}]{Papaloizou1995a}
{Papaloizou}, J.~C.~B., \& {Lin}, D.~N.~C. 1995, \apj, 438, 841,
  \dodoi{10.1086/175127}

\bibitem[{{Papaloizou} \& {Pringle}(1983)}]{Papaloizou1983}
{Papaloizou}, J.~C.~B., \& {Pringle}, J.~E. 1983, \mnras, 202, 1181,
  \dodoi{10.1093/mnras/202.4.1181}

\bibitem[{{Papaloizou} \& {Terquem}(1995)}]{Papaloizou1995b}
{Papaloizou}, J. C.~B., \& {Terquem}, C. 1995, \mnras, 274, 987,
  \dodoi{10.1093/mnras/274.4.987}

\bibitem[{{Pringle}(1992)}]{Pringle1992}
{Pringle}, J.~E. 1992, \mnras, 258, 811, \dodoi{10.1093/mnras/258.4.811}

\bibitem[{{Pringle}(1996)}]{Pringle1996}
---. 1996, \mnras, 281, 357, \dodoi{10.1093/mnras/281.1.357}

\bibitem[{Pringle(1997)}]{Pringle1997}
Pringle, J.~E. 1997, Monthly Notices of the Royal Astronomical Society, 292,
  136, \dodoi{10.1093/mnras/292.1.136}

\bibitem[{{Rakshit} {et~al.}(2020){Rakshit}, {Stalin}, \&
  {Kotilainen}}]{Rakshit2020}
{Rakshit}, S., {Stalin}, C.~S., \& {Kotilainen}, J. 2020, \apjs, 249, 17,
  \dodoi{10.3847/1538-4365/ab99c5}

\bibitem[{{Scheuer} \& {Feiler}(1996)}]{Scheuer1996}
{Scheuer}, P.~A.~G., \& {Feiler}, R. 1996, \mnras, 282, 291,
  \dodoi{10.1093/mnras/282.1.291}

\bibitem[{{Simon} {et~al.}(2012){Simon}, {Beckwith}, \& {Armitage}}]{Simon2012}
{Simon}, J.~B., {Beckwith}, K., \& {Armitage}, P.~J. 2012, \mnras, 422, 2685,
  \dodoi{10.1111/j.1365-2966.2012.20835.x}

\bibitem[{{Vaughan} {et~al.}(2016){Vaughan}, {Uttley}, {Markowitz},
  {Huppenkothen}, {Middleton}, {Alston}, {Scargle}, \& {Farr}}]{Vaughan2016}
{Vaughan}, S., {Uttley}, P., {Markowitz}, A.~G., {et~al.} 2016, \mnras, 461,
  3145, \dodoi{10.1093/mnras/stw1412}

\bibitem[{{Wijers} \& {Pringle}(1999)}]{Wijers1999}
{Wijers}, R. A.~M.~J., \& {Pringle}, J.~E. 1999, \mnras, 308, 207,
  \dodoi{10.1046/j.1365-8711.1999.02720.x}

\end{thebibliography}
\bibliographystyle{aasjournal}

\end{document}